# Recent advances and perspective of photonic bound states in the continuum

Guizhen Xu[†], Hongyang Xing[†], Zhanqiang Xue[†], Dan Lu, Jinying Fan, Junxing Fan[‡], Perry Ping Shum and Longqing Cong[*]

*Department of Electrical and Electronic Engineering, Southern University of Science and Technology, Shenzhen 518055, China*
[†]*These authors contributed equally to this work.* [‡]*fanjx@sustech.edu.cn.*

[*]Address correspondence to: conglq@sustech.edu.cn

**Abstract:** Recent advancements in photonic bound states in the continuum (BICs) have opened up exciting new possibilities for the design of optoelectronic devices with improved performance. In this perspective article, we provide an overview of recent progress in photonic BICs based on metamaterials and photonic crystals, focusing on both the underlying physics and their practical applications. The first part of this article introduces two different interpretations of BICs, based on far-field interference of multipoles and near-field analysis of topological charges. We then discuss recent research on manipulating the far-field radiation properties of BICs through the engineering of topological charges. The second part of the article summarizes recent developments in the applications of BICs, including chiral light and vortex beam generation, nonlinear optical frequency conversion, sensors, and nanolasers. Finally, we conclude with a discussion of the potential of photonic BICs to advance terahertz applications in areas such as generation and detection, modulation, sensing, and isolation. We believe that continued research in this area will lead to exciting new advancements in optoelectronics, particularly in the field of terahertz devices.

**Keywords:** Bound states in the continuum; Metamaterial; Photonic crystal; Terahertz device; Applications.

## 1. Introduction

Bound states in the continuum (BICs) are a ubiquitous wave phenomenon in various areas such as electromagnetism, acoustics, and fluid mechanics.[1] As trapped states with embedded eigenvalues, BICs exhibit intriguing localized energy in an open resonator or non-Hermitian system, even though the system is coupled to radiation continuum.[2] The first study of BIC dates back to 1929 when von Neumann and Wigner mathematically proposed an artificial quantum potential to support a BIC.[3] In 1985, BIC was found in a two-resonator system by continuous parameter tuning where destructive interference of the two resonances leads to vanishing linewidth in one of the resonances.[4] This type of BICs is known as Friedrich-Wintgen (FW) BICs or accidental BICs. The physics of BICs was not introduced to optics until 2008,[5] and then experimentally demonstrated with an array of coupled optical waveguides in 2011 where an antisymmetric mode was shown to propagate without loss to the continuum.[6] This type of BICs was known as symmetry-protected BICs that are decoupled to the continuum due to symmetry mismatch. Two years later, optical BICs were experimentally observed in a two-dimensional periodic photonic lattice – photonic crystal (PhC), by researchers from Massachusetts Institute of Technology, which initiates the deeper exploration of BICs in PhCs and metamaterials.[2]

Theoretically, BICs possess infinite radiative lifetimes in an infinitely periodic lattice enabling boundless enhancement of electric and magnetic fields. However, the finite extent of periodic lattices, intrinsic material absorption, fabrication defects, and structural disorders would result in a collapse of the divergent quality factor ($Q$). A leaky mode termed as "quasi-BIC" is commonly adopted for photonic applications.[7] Since ideal BICs only exist in theory and cannot be captured in far field in practice, BICs and quasi-BICs usually refer to similar leaky scenarios in the literature. It is important to note that the measured $Q$ from a Fano resonance enabled by BIC refers to the total quality factors ($Q^{tot}$) contributed by radiative $Q^{rad}$ and nonradiative $Q^{nrad}$, and it is thus crucial to reduce nonradiative losses for a larger $Q^{tot}$.

Bound states in the continuum (BICs) are a fascinating phenomenon in photonic crystals and metasurfaces. In this perspective article, we focus on two types of BICs: symmetry-protected and accidental BICs. They can be observed in a band above light cone at high symmetry points and nonzero wavevectors (Fig. 1a).[6, 8, 9] Fundamentally, the origin of BICs is explained by destructive interference. Specifically, diffraction orders form the discretized radiation channels to a finite number of directions in a periodic lattice, and only a single open diffraction channel remains when the lattice is constructed with subwavelength unit cells.[7] BICs are formed when the coupling of the single channel to free space is ceased due to symmetry mismatch (symmetry-protected BICs) or by continuous parameter tuning (accidental BICs).

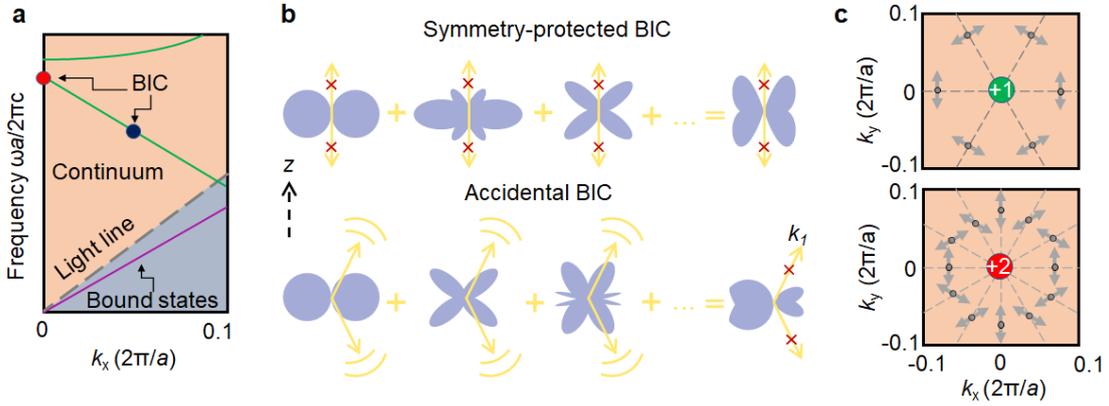

**Figure. 1. Interpretation of BICs from multipolar analysis and topological charges.** (**a**) Schematic of a typical dispersion diagram of photonic crystals. (**b**) BIC formation in terms of far field multipolar analysis. (**c**) Schematic of topological charges characterized by polarization vector in the momentum space.

The physical interpretation of BICs in periodic lattices can be categorized into two directions: multipolar analysis and topological defects.[10, 11] Multipolar analysis refers to the associated radiation patterns of multipoles as illustrated in top panel of Fig. 1b, where any leakage of energy is forbidden in the normal direction if all the multipoles are not allowed to radiate along the z-axis, leading to a symmetry-protected BIC at the Γ point in the momentum space.[10] In contrast, accidental BICs can be formed when all the contributed multipoles are real and interfere destructively in the direction of $k_1$, resulting in complete cancellation of radiation along the $k_1$ direction of the open diffraction channel.

From the perspective of topological defects, BICs were accurately manipulated in the platform of PhCs.[11, 12] As a typical open photonic system, PhCs have demonstrated a variety of applications in laser cavities,[13-18] waveguides,[19] and sensors[20-22] whose physical frameworks are beyond Hermitian systems. One unique feature in the non-Hermitian systems is BIC which has no counterparts in the Hermitian systems.[23] From the polarization vector fields in momentum space, BICs are considered as a topological defect whose polarization vector is undefined. Topological defects could be numerically characterized by a topological charge ($q$) representing the nontrivial winding patterns of certain parameters, e.g., velocity, phase, or polarization in real space.[11, 24] Specifically, for BICs in non-Hermitian systems, the topological charge is characterized by the winding of polarization vectors in momentum space as

$$q = \frac{1}{2}\oint_C dk \cdot \nabla_k \theta(k).$$

Here $\theta(k)$ indicates the angle of major axes of the polarization vectors in momentum space that forms mode of the radiation field $\psi(k) = arg\left[e_x\hat{x} + ie_y\hat{y}\right]$, and $C$ indicates a closed integration path to circle the state in the counterclockwise direction. According to the polarization vectors of radiation fields, a BIC renders a polarization singularity in the momentum space attributed to cancellation of radiation.[11] Since the polarization vector is not continuously defined, a nontrivial topological invariant $q \in \mathbb{Z}$ appears in analogy to a Chern number characterizing the nontrivial bulk band in a topological insulator. As a conserved and quantized quantity, topological charges are merely allowed to continuously evolve in the momentum space and remain constant unless the charge moves out of light cone or annihilates with an opposite sign charge. Such an interpretation of BICs has been experimentally demonstrated by directly observing the polarization vortices in the vicinity of mode center,[25-27] and found promising applications in engineering the robustness of BICs, vortex lasers and chiral BICs (refer to section III).

## 2. Engineering BICs from topological charges

### a. Merging BICs

Topological charges carry important characteristics of radiation, and the manipulation of them enables a flexible approach to tailor the radiation properties. Robustness is an important hallmark in the topological systems, in which the invariability is commonly prescribed for commensurate topological invariants of photonic bands mapping to momentum space such as Zak phase, Chern number, and winding number. [28-31] In this section, we will discuss recent advance in enhancing the robustness of BICs by manipulating their topological charges.

By virtue of the topological properties, a square lattice PhC slab was proposed to investigate the topological polarization singularity in momentum space.[24] In this work, a band with a BIC located at the Γ point which is guaranteed by the $C_2$ symmetry, as well as eight accidental BICs located at off-Γ points, was studied. The polarization vectors were calculated to carry a topological charge of +1 or -1 at the nine BICs, which exhibited diverging $Q^{rad}$ (Fig. 2a). In the particular band studied, a BIC protected by symmetry was fixed at the center of the Brillouin zone (BZ), while the remaining eight accidental BICs underwent evolution along the highly symmetric lines by continuously adjusting parameters such as the lattice period. All the BICs existed in the band protected by the nontrivial topological charges, and remained stable until the charges merged and annihilated into a single isolated BIC with a charge of +1 at the center of the BZ.

This merging process alters the intrinsic radiation behavior of an isolated BIC governed by the relation of $Q^{rad} \propto 1/k^2$, and additionally safeguards the radiation against scattering out of the cavity confined by the nearby accidental BICs (Fig. 2a). The topological property triggered a merging BIC, which complied with $Q^{rad} \propto 1/k^6$ and is very promising in suppressing the unexpected scattering losses due to fabrication defects or disorders, and improving the upper limit of $Q$ in practice. A record high $Q^{tot}$ of $4.9\times10^5$ was experimentally achieved by merging BICs, which was 12 times higher than the value of an isolated BIC. This milestone work could serve as the backbone in merging BICs to diverse more attractive studies.

To further explore the role of topological charges, high-order charges were studied in a hexagonal lattice PhC,[32] which exhibited merging BICs at the Γ point with a topological charge of -2, as illustrated in Fig. 2b. Moreover, a relation of merging BICs with a topological charge $q=\pm n$ could be generalized as $Q^{rad} \propto 1/k^{-2n-4}$. Another study demonstrated the confinement of light in all three dimensions by utilizing merging BICs that provided vertical confinement and photonic bandgap as mirrors for lateral confinement (Fig. 2c),[33] which has further improved the $Q^{tot}$ to a higher value of $1.09\times10^6$. The lateral confinement also ensured a localized Bloch mode surrounded by the bandgap PhC, leading to a reduced modal volume of $V = 17.74(\lambda_0/n)^3$. The simultaneous increase in $Q^{tot}$ and reduction of modal volume is expected to significantly enhance the Purcell effect, which would be beneficial for low-threshold lasers, chemical and biosensors, and nonlinear photonics.[17]

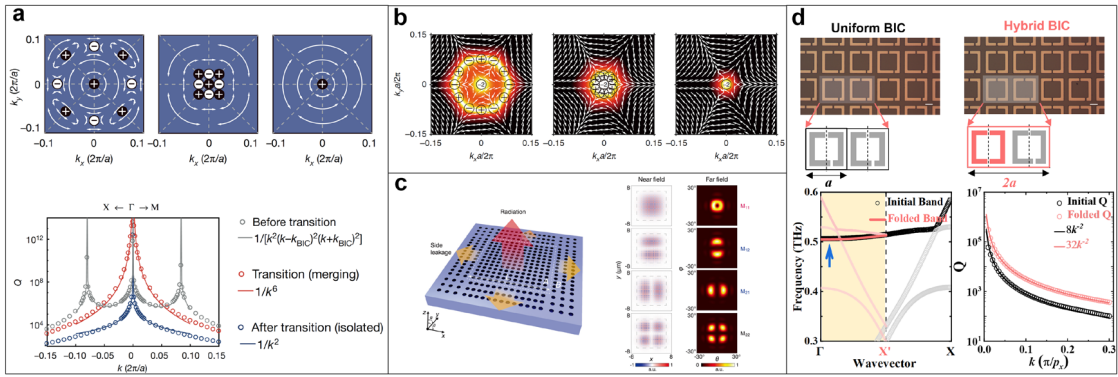

**Figure 2. Merging BICs and folded BICs to improve the quality factors in practice.**
(**a**) Dynamics of merging BIC in terms of topological charges by parameter tuning.[24]
(**b**) Merging BICs with high-order topological charges.[32] (**c**) Miniaturized BICs with a heterostructure to confine photons in all the three dimensions.[33] (**d**) Hybrid BIC metasurfaces with mixed nonradiative BIC and quasi-BIC resonators.

The concept of merging BICs effectively increases the slope of $Q^{rad}$ versus $k$ (see bottom panel of Fig. 2a), which mitigates the impact of fabrication defects and disorders on $Q^{rad}$ in practice. However, merging BICs have a stringent requirement of special band with multiple BICs in the momentum space and demand an extremely accurate design of resonators at sub-nanometer scale for the assembly of all the charges in the vicinity of Γ point. Recently, a generalized path was proposed to improve the lifetime and robustness of BICs with a hybrid BIC lattice. Different from conventional approaches to access quasi-BIC by breaking symmetry of resonators uniformly in a metamaterial lattice, the ability to selectively close radiative channels in the lattice as a whole allows for fine-tuning of the radiation density in the BIC and quasi-BIC hybrid lattice, as demonstrated in Fig. 2d. This is a generalized approach that could be extended to any symmetry-protected BIC without requirements of accurate geometric design or band selectivity. The hybrid BIC lattice offers a strong and reliable way to achieve high $Q$ resonances that are more than 14.6 times greater than the conventional lattices. Such an advanced structure provides an excellent method for obtaining precise, high-$Q$ resonances that are not affected by fabrication disorders, ultimately reducing the negative impact of common fabrication flaws on resonance quality. The process of band folding reveals the origin of the robust and high-$Q$ resonances in the hybrid BIC lattice and uncovers multiple Fano resonances that arise from the otherwise bounded eigenstates at X, Y, and M points in the momentum space. The robust and high-$Q$ multiple Fano resonances are especially important in the terahertz (THz) regime for pulse generation, sensing, and communications.

## b. BICs of splitting topological charges

To study circularly polarized states stemming from BICs, the in-plane $C_2$ symmetry was first broken, leading to pairs of circularly polarized states (*C* points) generated from the symmetry-protected BICs (*V* points as vortex polarization singularities, as shown in Fig. 3a).[26] A theoretical demonstration soon followed, showing that breaking the in-plane $C_2$ symmetry was not necessary for the emergence of *C* points, which could also appear near the K point and Dirac-degenerate points in a PhC with honeycomb lattice, with in-plane inversion symmetry and time-reversal symmetry conserved (Fig. 3b).[34] In addition, the work pointed out that only a pair of *C* points with identical (opposite) handedness and opposite (identical) topological charges could be annihilated (merged) together. However, this method is not universal since the achievement of *C* points is reliant on the incompatible symmetry of BICs (a double-degenerate BIC). Generation and annihilation of BICs and *C* points were also probed with higher order topological charges (-2) by breaking the $C_6$ symmetry. Two merging processes of *C* points were observed by breaking the up-down symmetry, where unidirectional radiative resonances with leaky channels of drastically different radiative lifetimes emerge.[35] Without breaking the up-down symmetry, the *C* points would not appear at Γ points, and circularly polarized radiation could not be measured at normal incidence. However, they could still be accessed at other points in the band, which is known as extrinsic chirality.[36, 37]

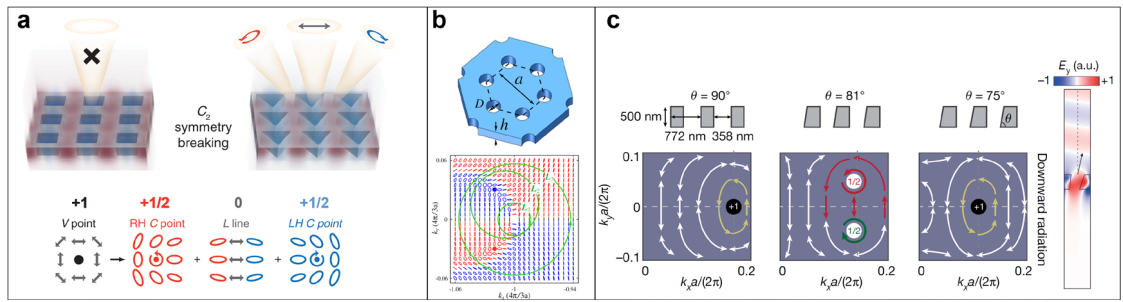

**Figure 3. Polarization manipulation and asymmetric radiation enabled by topological charge spitting.** (**a**) Circular polarization radiation at Γ point by splitting topological charge (+1) in the process of symmetry breaking of a PhC with intrinsic

chirality, and (b) circular polarization radiation near K point and Dirac-degenerate points in a honeycomb lattice PhC.[26, 34] (c) Unidirectional radiation by splitting topological charge with slant geometry by breaking up-down symmetry (z-axis).[38]

When the up-down symmetry was broken, the first experimental demonstration of unidirectional radiation was realized in a one-dimensional silicon PhC slab with slant sidewalls as shown in Fig. 3c. A TE-like band was found to have an accidental BIC at off-Γ point in the reciprocal space that possesses $q = +1$ when the in-plane mirror symmetry and out-of-plane (z-axis) symmetry is conserved.[38] When the out-of-plane symmetry was preserved, the radiation properties were the same along the +z and -z directions. However, when the symmetry was broken by tilting the sidewalls of the grating along the z-direction, the radiation symmetry was also broken, resulting in the splitting of the +1 charge into paired half-integer charges. These charges were protected by the symmetry along the y-axis (Fig. 3c). At the same time, BIC disappeared and the radiation decay rates towards +z and -z directions decoupled for the two half-integer charges whose radiation was circularly polarized with opposite helicities. By continuously tuning the tilting angle of the slant sidewalls, the two half-integer charges merged into an integer charge for the downward radiation forming an accidental BIC again, but the radiation channel along the upward direction was still open. This specific profile enabled an ideal asymmetric unidirectional emission with measured asymmetric ratio as high as 27.7 dB and $Q^{tot}$ of $1.6 \times 10^5$.

## c. Emerging mechanisms and applications of BICs

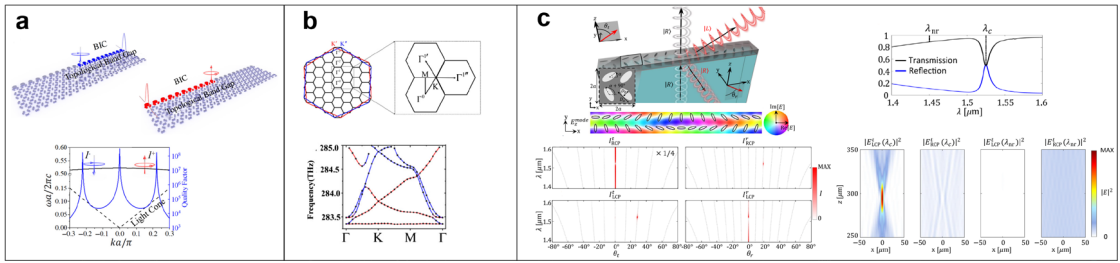

**Figure 4. BICs enabled by different physics and their applications. (a)** Bound
9

topological edge states in the continuum.[39] (**b**) Flat band BICs.[40] (**c**) Application of narrow band BICs in non-local metasurfaces.[41]

Radiation is a ubiquitous phenomenon to bring non-Hermicity to photonic systems, so that a topologically robust high-$Q$ resonance paves the way to understand the intrinsic physics underneath the general system. The higher $Q$ led to a promising platform to enhance light-matter interactions in topological photonics, which is beneficial to polish the performance in topologically integrated optoelectronic devices.[42] Merging BICs also provide alternative methods for realizing large-area lasers combined with optical cavities, modulating light radiation using active media, and generating new frequency components through nonlinear effects. Recently, interesting mechanisms of BICs integrating with topological photonics and flat band were reported, and applications in wavefront manipulation were introduced with nonlocal metasurfaces.

**Topological and flat band BICs.** One of the most canonical behaviors in topological photonics is the support of the topologically protected edge states confined on the interface between topologically trivial and nontrivial PhCs.[43, 44] In 2015, an all-dielectric photonic system was proposed to exhibit a pair of spin-locking edge states within the bandgap analogous to quantum spin Hall effect.[45] This type of edge states mainly exists in the vicinity of Γ point and above the light cone, which provides a potential condition to investigate the interactions between BICs and edge states. In addition to the bound topological edge states in the continuum explored in 2021 (Fig. 4a),[39] another method to combine BICs with topological flat bands was proposed by extending the flat band enabled by bilayer honeycomb PhC at a magic angle into the continuum as shown in Fig. 4b.[40] The band with a BIC at Γ point could be folded and forms a flat band under the appropriate twist angles and spatial distance of two identical PhC slabs. This work provides an adjustable way to trigger a high-$Q$ quasi-BIC by introducing one more flexible degree of freedom, removing the requirements of geometric symmetry-broken or the precise incident angles. To explore greater dimensions, Floquet edge states[46, 47] and corner states in high-order topological insulators[48-50] were also studied to analyze the connection between topology and BICs.

Topological BICs not only inherit the intrinsic properties originated from dispersion relationship in topological systems, but also provide additional radiation properties towards free space. Topological BICs paved a feasible way to couple and radiate topological edge states to free space, which omits the complicated coupling channels in topologically integrated chips. In addition, flat band and flexibly adjustable $Q$ are beneficial to second harmonic generation in nonlinear materials on which the incident angle of pump light exerts a negligible impact. To date, topological photonics has combined with nonlinearity,[51, 52] non-Hermicity,[53, 54] and synthetic dimensions[55] to explore topological properties in multiple degrees of freedom, and thus exhibits a plethora of scenarios to investigate the physics of BICs. The combination of topologically protected edge states and BICs guaranteed both lateral and vertical confinement and would lead to more tight mode localization with further reduced radiative losses.[33] Another interesting combination lies at exploring BIC-supported zero-index media in PhCs by designing a Dirac cone at Γ point in the band structure.[56, 57] Such a combination could enable longer coherence length with low losses and benefit linear and nonlinear optics in a large-area medium.

**Applications of BIC in nonlocal metasurfaces.** Manipulation of wavefront can be achieved through metasurfaces, which can be classified into two types: local and nonlocal metasurfaces. Local metasurfaces engineer the wavefront by individually manipulating each unit cell,[58-60] whereas nonlocal metasurfaces support resonant modes spatially extending across adjacent unit cells which are formed by collective excitation.[61-63] BICs and quasi-BICs (manifesting as Fano resonances) are typically induced in nonlocal metasurfaces. Although local metasurfaces have revealed extraordinary capability to shape the wavefront, they commonly lack spectral selectivity.[58, 64-66] In contrast, nonlocal metasurfaces offer great spectral selectivity with a narrow linewidth.[67, 68] In the past few years, nonlocal metasurfaces has been widely applied in sensing,[20, 21] optical signal processing,[68] augmented reality,[69] and harmonics generation.[70, 71] However, traditional nonlocal metasurfaces have limited capabilities in tailoring wavefront.

Simultaneous manipulation of light in both spatial and spectral domains was realized with nonlocal metasurfaces supporting BICs, as shown in Fig. 4c.[41, 72] Symmetry-protected BICs were utilized to achieve frequency selectivity and introduce a geometric phase that operates only within the spectral range of the quasi-BIC resonances. The Pancharatnam-Berry (P-B) phase enables additional phase of 4α (α is orientation angle of dimers) for the transmitted left-handed circular polarization (LCP) and reflected right-handed circular polarization (RCP) with RCP incidence, while the phases of reflected LCP and transmitted RCP are invariant to α. Therefore, any phase profile is feasible to engineer the wavefront in the spectral vicinity of quasi-BICs. Note that the electric field of nonlocal metasurface exhibited a collective supermode (Fig. 4c), and the nearest-neighbor coupling led to a blueshift of resonance from the eigenvalue of isolated resonators in the nonlocal metasurfaces. Consequently, the deflection only occurred at a narrow wavelength range of BICs leaving the remaining spectrum transparent without wavefront distortion. Nonlocal metasurfaces also enable approaches to control more modes simultaneously.[63]

Light focusing of nonlocal metasurfaces to function as metalenses is feasible with larger numerical aperture (NA) as well as narrow linewidth. For nonlocal metalenses, NA can be described by: $NA^2 \leq \frac{\omega_0 / k_0^2}{|b| Q}$, where $b$ is a constant, $Q$ is quality factor, $k_0$ is the free space wave vector, and $\omega_0$ is the angular frequency of the mode at $k = 0$. A flat band ($|b|$ is small) would compensate the high-$Q$ resonance and offer a high NA.[41, 73] Based on the idea of nonlocal metasurfaces, researchers have demonstrated the creation of multifunctional metalenses that can modulate resonant wavelength, $Q$ factor, dispersion, and wavefront at multiple wavelengths. This was achieved by cascading nonlocal metasurfaces, while maintaining transparency for all other wavelengths.[74] The nonlocal BIC metasurfaces hold great promise for augmented reality (AR) glasses as a beam splitter with the capability to control wavefront at selected frequencies (i.e., colors) while remaining transparent at off-resonance frequencies.[69, 75]

In addition to the full control of wavefront of coherent light, partially coherent radiation from thermally emitted light could be manipulated by nonlocal BIC metasurfaces.[76] The thermal metasurfaces that could radiate with selected wavefronts, controllable linewidth, spin and angular orbital moments, and coherence were demonstrated. Nonlocal BIC metasurfaces would enrich applications in wavefront selection,[77] quantum optics,[78] and nonlinear optics.[79, 80] It should be noted that efficiency remains a common challenge for both local and nonlocal metasurfaces, with a theoretical limitation of 25%. However, chiral BIC metasurfaces (see section III) have shown the potential to overcome this limitation, achieving unlimited efficiency of up to 100%.[81]

3. Applications

a. Chiral light and vortex beam generation

Chirality, as a fundamental feature of nature, refers to the geometric attribute of objects that does not coincide with their mirror images.[82, 83] Metamaterials composed of chiral elements exhibit strong chiroptical responses,[84] which have been demonstrated by metamaterials with helices,[85, 86] N-shaped structures,[87-89] and multi-layered structures.[90-92] However, achieving efficient and high-$Q$ chiroptical effects remains a challenge, which leads to the limited ability in chiral emission,[93] circular polarization-sensitive photodetection,[94] and chiral molecular separation[95] due to weak chiral light-matter interactions.

Consistent with the generation of circularly polarized states by topological charge splitting, intrinsic chirality was explored under BIC schemes. Maximal chirality could be reached by utilizing dimers composed of classical dielectric elliptical bars with symmetry-protected BICs, which could be numerically analyzed with coupled-mode theory (Fig. 5a).[96] Achieving optimal chirality requires precise control over the radiation coupling of eigenstates to free space , and BICs provide an ideal starting point for analysis as they are fully decoupled from free space at their initial state. By breaking $C_2$ symmetry with an increase of $\theta$, BICs were transformed into quasi-BICs, but the overall structure remained intrinsically achiral with merely in-plane symmetry broken. Out-of-plane symmetry was further broken by introducing a vertical offset ($d$) of the elliptical bars, which eliminated the mirror symmetry, and thus intrinsic chirality emerged. By adjusting the parameters $\theta$ and $d$, the dissipation matching condition could be satisfied in the critical coupling regime, allowing for the achievement of maximal chirality. In addition to breaking symmetry by tuning parameters $\theta$ and $d$, chiral quasi-BICs were also systematically studied with double-layer metasurfaces,[81] and maximal chirality could be achieved by adjusting the orientation angles ($\alpha$) of elliptical bars and the relative rotation ($\Delta\alpha$) between the up-down bars in the two layers of metasurfaces (Fig. 5b). Additionally, the adjustable parameters provide freedom to freely control the radiation phase and guaranteed maximal transmission amplitude. BICs enable highly efficient wavefront shaping of circularly polarized light and sharp frequency selectivity. The utilization of BICs provides a promising approach to surpass the traditional 25% efficiency restriction in conventional planar metasurfaces under linear polarization illumination.[41] However, breaking the up-down symmetry of intrinsic chirality required to implement these designs is currently only possible in the microwave range, where three-dimensional samples are more easily accessible.[97] Implementing these designs in the infrared and visible ranges remains challenging due to fabrication difficulties. One solution was reported with PhCs of slant sidewalls breaking both in-plane and out-of-plane symmetries at visible frequencies,[98] which enabled that chiral emission with giant field enhancement and a small divergent angles were experimentally observed (Fig. 5c).[93] Although fabrication in nanoscale is challenging, multi-step fabrication solutions were proposed for dielectric dimers with arbitrary height difference.[99] This approach would remove the fabrication barrier of breaking the up-down symmetry in PhCs, enabling the experimental exploration of more intriguing phenomena in shorter wavelength. Most recently, planar dielectric metasurfaces were reported with optimized efficiency of circularly polarization conversion and very narrow spectral linewidth empowered by BICs without breaking out-of-plane symmetries.[100] Similar to chirality, the absorption of circularly polarized light was studied on the basis of BICs with a stereoscopic split ring resonator array covered by metals on the surface in THz regime.[101]

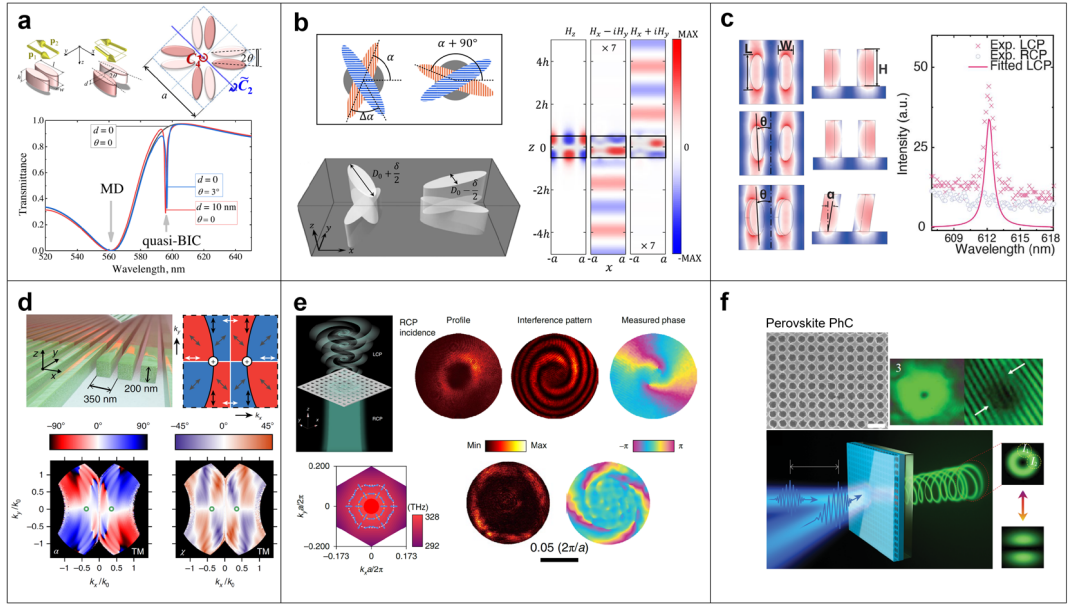

**Figure 5. Applications of BICs in intrinsic chirality and vortex beam generation.** (**a**) Chiral BIC metasurfaces with a maximal circular dichroism.[96] (**b**) Chiral BIC metasurfaces supporting high efficiency wavefront shaping.[81] (**c**) Chiral emission from a BIC metasurface with slant sidewalls.[93] (**d**) Experimental observation of polarization vortex from BICs.[25] (**e**) Vortex beams generation via momentum-space polarization vortices centered at a BIC.[102] (**f**) Vortex lasing based on a perovskite PhC.[14]

One more intriguing aspect of BICs is their ability of light manipulation through spin Hall effect with circularly polarized light.[103] The strong spin-orbit interactions are enhanced by the topological vortices around BICs, where wavevector dependent P-B phase gradient and conventional P-B phase gradient of cross-polarized light lead to the synthesized spin-dependent lateral light shifts. Vortex beam is another significant consequence of topological charges of BICs in radiation.[25] Early in 2011, vortex beams with quantum numbers of optical angular moment $l$ =1, 2, and 3 were demonstrated using PhCs at Γ point, which possessed topological charges of $q$ = 1, 2, and 3.[104] Interpreted from BICs, it is anticipated that the light radiated by modes around BICs would display a vortex in the far-field polarization profile. The first experimental observation of full polarization states around a BIC was reported in ref.[25] demonstrating the presence of topological vortex as shown in Fig. 5d. Characterized by polarization ellipse, the vortex exists not only in polarization angle ($α$) but also in ellipticity ($χ$). Similar to the multipolar analysis, BICs in the grating can be explained as a Friedrich–Wintgen bound state arising from the interference between two resonant dipolar contributions ($z$-oriented electric and $y$-oriented magnetic dipoles). Based on the strong polarization anisotropy in the vicinity of BICs in momentum space of a PhC, P-B phase will be added to cross-polarized radiation with a circularly polarized incidence. Different $k$ components of incidence would interact with the $k_{∥}$ resonances, and the transmitted cross-polarized radiation would automatically possess a spiral phase front with an optical angular momentum (topological charge) $l = \mp 2 \times q$ where $q$ is the polarization charge of BICs and $\mp$ is for RCP and LCP incidence, respectively (Fig. 5e).[102] The generated vortex is a high-order quasi-Bessel beam that is diffraction resistant, and the PhC has no structural center, making it free from alignment. Higher orders of orbital angular momentum (OAM) can be easily accessed by using samples of higher symmetry or by tuning the working wavelength. Beyond passive devices, the generation of vortex beam in radiation is also feasible in cavities with active media. Under the similar interpretation of topological charge of BICs, vortex lasing was reported in a microlaser based on a lead bromide perovskite PhC (Fig. 5f).[14] Furthermore, BIC condition could be broken by introducing imaginary part in the cavity in an ultrafast scale (1-1.5 ps), and thus ultrafast switching of the polarization from vortex to linearly polarized beam was demonstrated with ultralow energy consumption.

**b. Nonlinear optical frequency conversion**

BICs have received great attention in the field of nonlinear frequency conversion. A plethora of interesting PhCs and metamaterials based on BICs or quasi-BICs have opened up a new path to boost harmonics generation with high yield and efficiency owing to high-$Q$ and extreme local field enhancement. Nonlinear optical processes at nanoscales, such as second-, third- and high-harmonic generation, THz generation, and frequency mixing effects, have attracted significant interest in both scientific and industrial development for high conversion efficiency and miniaturized devices. Regarding nonlinear effects in conventional bulky media, optical pump of high intensity is typically required to observe harmonics given the inherently weak nonlinear interactions, and complex phase matching between the fundamental and harmonics is essential. By contrast, metamaterials and PhCs relax the strict phase-matching conditions and enhance local fields in a tight volume.[105, 106] In this section, we review recent advance in nonlinear optical frequency conversion with BIC metasurfaces.

**Second harmonic generation.** Second harmonic generation (SHG) relies on the second order nonlinear susceptibility of crystals, which only occurs in non-centrosymmetric crystalline structures, and thus the studies mostly concentrated on III-V semiconductors materials with high second-order coefficients, such as AlGaAs,[70, 107-110] GaP[111, 112] and GaAs.[113, 114] Recently, an isolated AlGaAs resonator was reported to support a quasi-BIC interpreted from multipolar analysis, and further reduction of radiative losses was realized by optimizing the substrate configuration with an epsilon-near-zero (ENZ) layer (Fig. 6a). The high-$Q$ quasi-BIC enabled an experimental observation of SHG efficiency as high as $1.3\times10^{-6}$ W$^{-1}$, which was more than 2 orders of magnitude higher than other types of resonances.[70] With a metasurface array of AlGaAs resonators, SHG efficiency could theoretically be boosted up to 10% at a moderate pump intensity of 5 MW/cm$^2$ with a symmetry-protected quasi-BIC that exhibits 6 orders of magnitude higher than magnetic dipole resonances.[110]

Early in 2018, Fano resonances rooting in symmetry-protected BICs revealed a great improvement of SHG with a symmetry-broken L-shaped GaAs metasurface compared with the symmetry-conserved nanodisk GaAs slabs.[114] Metasurfaces composed of dimers of elliptical cylinders were also probed with symmetry-protected BICs to boost the SHG on a GaP material platform, which offered a relatively high nonlinearity, high refractive index, and low loss. The SHG efficiency of 0.1% W$^{-1}$ was achieved with 100-times lower intensities of pulsed pump (10 MW/cm$^2$) compared with previous reports in the literature. Additionally, efficient SHG was demonstrated on this platform with a conversion efficiency of $5\times10^{-5}$ % W$^{-1}$ pumped at the central wavelength of the quasi-BIC resonance with a continuous wave intensity of 1 kW/cm$^2$ (Fig. 6b).[111]

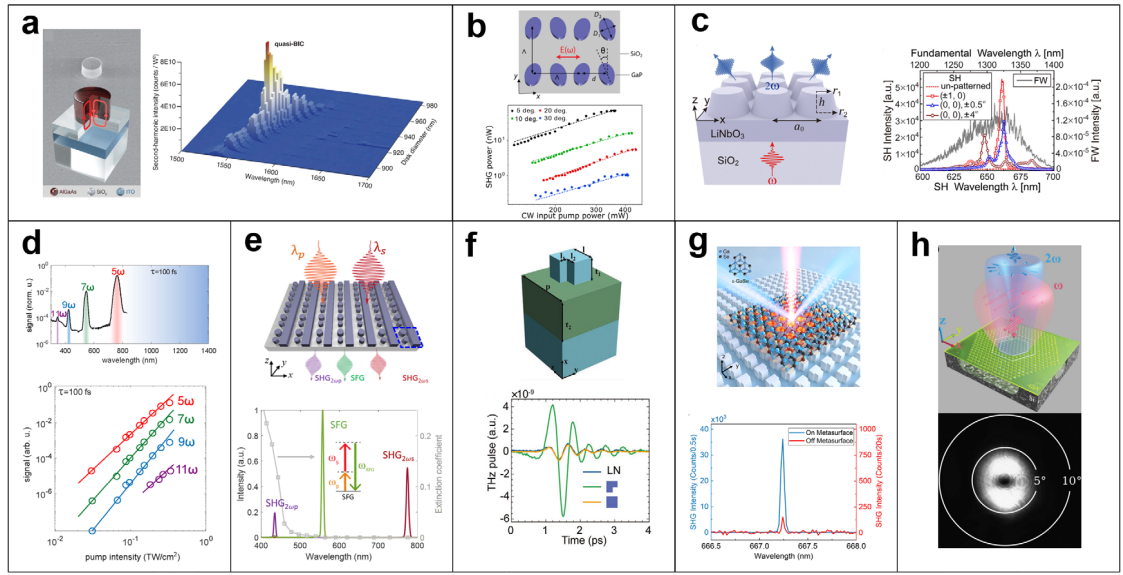

**Figure 6. Applications of BIC in nonlinear metasurfaces.** Efficient SHG in (**a**) an isolated AlGaAs resonator,[70] (**b**) GaP[111] and (**c**) LN metasurfaces supporting BICs.[115] (**d**) THG and HHG in an isolated silicon nanoparticle supporting BICs.[116] (**e**) SFG in GaP metasurfaces boosted by BICs.[117] (**f**) Single-cycle THz generation in LN metasurfaces enhanced by BICs.[118] (**g**) SHG enhancement in a hybrid BIC metasurface integrated with a few-layer GaSe.[119] (**h**) Vortex beam of SHG in a GaN PhC empowered by BICs.[120]

Low-index materials, such as lithium niobate (LN), have large second-order nonlinear susceptibilities and transparent window from ultraviolet to infrared,[121] which are promising for SHG despite of the difficulty in accurate fabrication. Although deep etch of LN with an accurate geometry is challenging, shallow etch would enable flexible engineering of the abundant modes in PhCs and induce high-$Q$ BICs. Initiated from the guided resonances in the unpatterned LN slab, an accidental BIC was observed by the interference between TE and TM modes in the shallow-etched LN metasurface which would significantly boost the SHG emission.[122] Experimental demonstration of enhanced SHG boosted by symmetry-protected BICs was reported in ref. [115] as shown in Fig. 6c. Shallow-etched *z*-cut LN metasurfaces were probed to utilize the largest nonlinear susceptibility, and SHG was enhanced by TM modes. More recently, hexagonal boron nitride (hBN) metasurfaces made of low-index materials were used for SHG covering the whole visible spectrum, hosting symmetry-broken BICs with an enhancement factor above $10^2$.[123] SHG is commonly probed in materials without centrosymmetry, and second-order nonlinear optical responses are not allowed in the bulk with centrosymmetry such as silicon and germanium. Nevertheless, the surface has no centrosymmetry and is accessible to SHG. Taking advantage of BICs to strengthen the local fields, efficient SHG in a symmetry-broken silicon metasurface was experimentally observed with an efficiency of $1.8\times10^{-4}$ $W^{-1}$.[124]

**High order harmonic generation and frequency mixing.** Third harmonic generation (THG) and high order harmonic generation (HHG) at nanoscales with quasi-BICs have mainly concentrated on silicon-based nanostructures.[71, 116, 125-128] Based on quasi-BICs in an isolated silicon nanoparticle similar to ref. [70], high-order harmonics including the third and fifth harmonics were obtained.[125] In a symmetry-protected BIC metasurface in the mid-IR regime, odd harmonics up to the 11$^{th}$ order were observed, and the transition from perturbative to nonperturbative regimes was traced as shown Fig. 6d.[116]

Moreover, there are also intricate frequency mixing phenomena at the nanoscales that involve at least two frequencies, such as sum-frequency generation (SFG), four wave mixing (FWM) and difference frequency generation (DFG). Highly efficient SFG was probed with high-*Q* metasurfaces empowered by BICs in the material platform of GaP (Fig. 6e).[117] Illuminated by pump beam at 875 nm and 1545 nm, a normalized conversion efficiency of $2.5\times10^{-4}$ $W^{-1}$ was obtained for SFG in the visible. Different from harmonic generation, the maximal efficiency of SFG was measured for inputs of nonparallel polarization. Similarly, DFG could also be enhanced by BICs and the resultant longer wavelength signals, e.g., THz emission, could enable compact light sources.[118, 129] With giant second-order sheet nonlinear susceptibility, single-cycle THz emission has been fully explored in metallic metasurfaces enhanced by magnetic-dipole resonances.[130-132] However, the large ohmic loss, low damage threshold and small interaction volume of plasmonic resonators hinder the nonlinear conversion efficiency in metallic resonators. Dielectric metasurfaces would avoid all the above-mentioned shortcomings, and efficient THz pulse emission boosted by BICs was recently reported based on LN platform. Localized field enhancement in LN film was obtained by the patterned $SiO_2$ metasurface due to the relatively lower index of $SiO_2$, and 17 times improvement of THz emission at 0.7 THz in contrast to the unpatterned LN thin film was experimentally observed as shown in Fig. 6f.[118]

**Hybrid and functional nonlinear metasurfaces.** Two-dimensional (2D) materials such as transition metal dichalcogenides and gallium monochalcogenides have revealed as excellent nonlinear optical materials with giant nonlinear susceptibilities. However, the short interaction length with light limits their nonlinear responses. The solution was raised by integrating the thin films with high-$Q$ resonators, and hybrid metasurfaces supporting BICs enable an excellent approach for enhancing interactions with 2D materials.[119, 133-137] Few-layer gallium selenide (GaSe) is a typical example that exhibits giant second order nonlinear susceptibility of ~1700 pmV$^{-1}$ at telecom band.[119] Boosted by silicon-based metasurfaces supporting symmetry-protected BICs, the hybrid GaSe-metasurface revealed a 9400-fold improvement of SHG compared with a bare GaSe flake (Fig. 6g).[119]

Functional nonlinear metasurfaces integrating harmonics generators and manipulators would further enrich the potential applications in which phase, amplitude, polarization and radiation direction of harmonics were finely controlled.[120, 138-141] Asymmetric radiation is an interesting topic both in linear and nonlinear regimes.[38, 138, 142] In nonlinear regime, asymmetry radiation of SHG in the forward and backward directions was theoretically proposed in a bianisotropic metasurface supporting quasi-BICs.[138] Highly normal-direction concentrated vortex beam was observed in the SHG based on a GaN PhC that supports doubly resonant BICs (Fig. 6h).[120] The great enhancement of SHG by the high-$Q$ resonances led to a conversion efficiency of $2.4\times10^{-2}$ $W^{-1}$ under continuous wave excitation. With the giant enhancement of harmonics generation, dynamical control of wavefront of harmonics was feasible via tuning polarization and wavelength.[140]

Boosted by BICs in metasurfaces, it is expected to enhance the nonlinear responses; however, accurate fabrication of nanostructures remains a challenge for certain materials despite the advanced fabrication technologies of silicon. Factors that need to be considered in the fabrication process include geometry, simplicity of fabrication, surface roughness, and others.[143] For interesting materials without appropriate fabrication approach, an alternative solution is to excite BICs in a metasurface with a low-index material so that the mode is largely confined in the high-index nonlinear materials to boost light-matter interactions.[144, 145]

c. Sensors

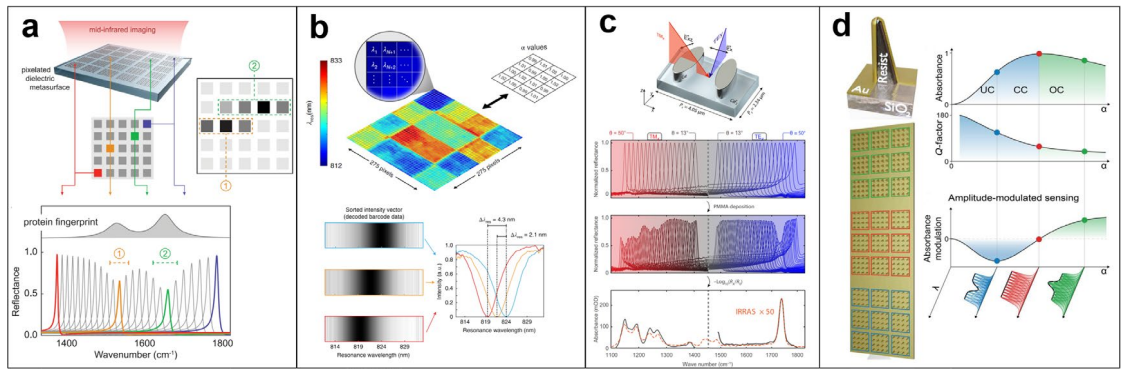

**Figure 7. Applications of BIC in sensors. (a)** Imaging-based molecular barcoding with a pixelated BIC metasurface.[20] (**b**) Hyperspectral imaging and biodetection enabled by BIC metasurfaces.[22] (**c**) Broadband molecular fingerprint retrieval with an angle-multiplexed BIC metasurface.[21] (**d**) A BIC enhanced senor with a plasmonic metasurface to further enhance local fields.[146]

With the goal of miniaturization and improved sensing performance, highly sensitive sensors that can be integrated onto a chip with ease of use, good portability, and free from maintenance, have become increasingly important.[1] The unique characteristics of symmetry-protected BICs, such as narrow linewidth, wide frequency scalability, and a well-defined spectrum over a broad wavelength range, provides a promising platform for sensing technology, particularly for imaging-based molecular barcoding in a pixelated configuration (Fig. 7a).[20] Absorption signatures of various molecules could be mapped in the spectra constituted by a series of quasi-BICs, each of which was linearly encoded to spatial location of the pixels. The BIC-based molecular barcoding provides more sensitive detection of molecular fingerprints with spectral resolution determined by linewidth of quasi-BICs. In this configuration, vibrational information of analyte could also be readout for chemical identification and compositional analysis. Similar BIC configuration was adopted to develop a one-shot imaging-based optical data acquisition and processing method, as shown in Fig. 7b.[22] An ultrasensitive label-free biosensing platform was demonstrated by combining the pixelated BIC metasurfaces with a commercial CMOS camera, capable of sensing less than three molecules per $\mu m^2$. The hyperspectral decoder approach correlated spectral information with spatial index of CMOS pixels, allowing high-resolution spectral information to be extracted from a one-shot image at a fixed wavelength. This is a wise spectrum-extraction scheme to eliminate the need for the bulky and expensive spectrum instrumentation. Integrating the pixelated metasurface chip with optofluidic system enabled more flexible and real-time detection of biomolecules. The reported sensitivity of 701/RIU and 0.41 nanoparticle/$\mu m^2$ for detecting a 204 femtomolar solution demonstrated the potential of this approach in detecting breast cancer extracellular vesicles encompassing exosomes. In addition to continuously tuning the central wavelength of symmetry-protected BICs in parameter space at normal incidence, similar modulation could also be accessible in momentum space by continuously tuning the incident angles (Fig. 7c).[21] The type of angle-multiplexed BIC spectra delivered a larger number of resonances (more than 200 resonances) from a single metasurface chip than those in geometry parameter space, which were limited only by the instrumental angular resolution (1.4 cm$^{-1}$ spectral step size) and range of incidence angles (13°-60°). Similar mechanism used in ref.[20] was applied to capture the molecular absorption fingerprint with high sensitivity (0.27 pg/mm$^2$) and chemical specificity.

A high-$Q$ resonance with narrow linewidth is essential for a good sensor to clearly resolve the spectrum caused by analyte, and the near-field interactions with the analytes can significantly influence overall sensitivity. Plasmonic sensors have been adopted to enhance the local field within the subwavelength scale, but they typically feature low-$Q$ resonances around 10 due to the intrinsic losses in metals.[146-148] Empowered by symmetry-protected BICs, a plasmonic nanofin metasurface was reported with $Q$ up to 180, as shown in Fig. 7d.[146] Quasi-BICs were realized by breaking the out-of-plane symmetry of the nanofins via introducing a slanted sidewall with an angle $\alpha$ fabricated by 3D laser nanoprinting, and the nanofin transformed from a rod to a triangular-type structure. Radiative losses could be tailored by adjusting the angle $\alpha$, and the coupling regime of the cavity could then be tuned to cross the undercoupling, critical coupling, and overcoupling regimes. In this work, a similar pixelated metasurface taking advantages of the broad wavelength coverage of BICs was developed for molecular sensors. Interestingly, different types of modulation in the spectra of the pixelated metasurface occurred, dominated by the coupling regimes of the resonators. Instead of metallic metasurfaces, applications of BICs in dielectric resonators would reduce Ohmic losses, further improving the overall $Q$ of resonances and thus a better sensitivity. A sensitivity of 178 nm/RIU was reported in a dielectric metasurface supporting a quasi-BIC with $Q$ of 2000 in the visible and infrared regimes.[149] Ultralow molecular weight detection of 186 Da was demonstrated with a 6 nm spectral shift per less than a 1 nm thick single molecular layer, showing the excellent sensing performance of BIC-based sensors. Sensors based on symmetry-protected BICs have also been reported in THz regime,[150] and applied to enhance the fluorescence emission and Raman scattering.[151]

### d. Nanolasers

The future of information processing is likely to involve photonic integrated circuits compatible of parallel signal processing, with nanolasers as the most critical integrated components.[13] Highly efficient room-temperature nanolasers that feature a compact cavity and broad wavelength selectivity are in high demand for various applications, including biological imaging, near-field sensing, integrated photonics, adjustable light sources, and cutting-edge display technologies.[152, 153] Recent developments of BIC physics enable new strategies to access long-lived and spatially confined resonances that offer precise manipulation of far-field lasing radiation.

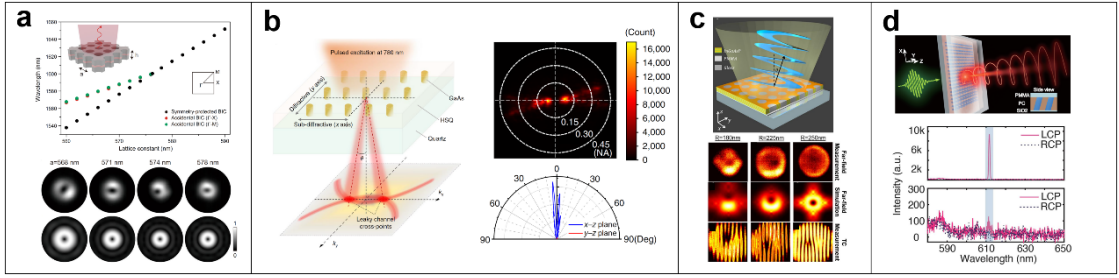

**Figure 8. Applications of BICs in lasing.** (**a**) A nanolaser with ultralow-threshold and reduced mode volume empowered by merging BICs.[17] The directivity of nanolasers controlled by altering (**b**) the period of the arrays[16] and (**c**) the symmetry of PhCs.[154] (**d**) Chiral lasing with left-handed circular polarization emission achieved by breaking up-down symmetry of PhCs.[93]

Early in 2017, a single-mode lasing action based on BICs in PhCs was reported at room temperature consisting of several $In_xGa_{1-x}As_yP_{1-y}$ multiple quantum wells that operated around the telecommunication wavelength[15]. The intrinsically infinite $Q$ of a BIC ensured that lasing could occur in arrays as small as 8×8 resonators. Typically, the smaller size of a BIC cavity leads to a rapid decrease in $Q$. However, merging BICs can slow down this trend, ensuring a high $Q$ resonance even in a cavity with smaller footprint. The idea was applied in a nanolaser based on an infinite-size InGaAsP PhC slab as shown in Fig. 8a.[17] Lasing threshold was found to be 1.47 kW/cm² that is much lower than other BIC-based lasers. Mode sizes also decreased at merging BIC states, and the far-field radiation patterns revealed a clear shrink until the merging state at $a$ = 574 nm.

A comprehensive investigation of lasing operation based on BIC metasurfaces was presented in ref.[18]. The results showcased the use of symmetry-protected BICs for low threshold lasing (1.25 nJ) with a high spontaneous emission coupling factor of 0.9, while maintaining a high $Q$ resonance. In addition to low-threshold to lasing, directional lasing is another key factor.[16, 30] Although subwavelength PhCs do not naturally permit radiation at BICs, it is possible to create controllable leaky channels by altering the period of the array, thereby adjusting the diffraction orders (Fig. 8b).[16] High $Q$ could be sustained even certain leaky channels were open, and excellent directivity with tunable lasing angles covering 0 to 25° could be achieved, as demonstrated in a GaAs nanopillar array.

With the ability to control lasing threshold and directivity, BIC mechanisms also offer the potential for manipulating lasing polarization and wavefront. For instance, BICs have been shown to generate vortex beams, which have unique characteristics of hollowness, diffraction-resistance, and quantized OAM. By integrating BICs with active media, e.g., InGaAsP multiple quantum wells, coherent vortex beams have been generated, with the lasing direction of the vortex beams controlled by adjusting the symmetry of resonators (Fig. 8c).[154] In addition, ultrafast control of lasing polarization between vortex and linearly polarized beams was demonstrated in a perovskite-based microlasers.[14] As discussed in the section of chiral light generation, another polarization state - circular polarization, can also be derived from BICs, and intrinsic chirality spawned from up-down symmetry breaking. Chiral quasi-BICs were demonstrated with high-$Q$,[93, 98] and chiral lasing with left-handed circular polarization emission was observed at 612 nm with a threshold of 22.14 mJ/cm$^2$ (Fig. 8d).

Lasing based on various types of hybrid PhCs has been reported, such as CdSe/CdZnS

core-shell nanoplatelets in the visible range.[155] Incorporating with high-$Q$ resonances of PhCs, dramatic spectral and divergence angle enhancement of lasing were demonstrated from a 100 nm thick organic molecule solution , resulting in significantly reduced threshold.[156] Meanwhile, BICs were characterized by their narrow spectral linewidths, making them highly sensitive to changes in refractive index and ideal for use as sensors. Nanolaser-based sensors not only offer coherent radiation, but also exhibit exceptional sensing capabilities.[157]

4. **Perspective on terahertz applications and conclusion**

THz radiation has exhibited great potentials in applications of ultrafast dynamics, biomedical diagnostics, security inspection, wireless communications and imaging.[158, 159,160] The capabilities of non-destructive imaging and label-free sensing become the unique features of THz waves due to the deep penetration, low photon energy, and abundant spectral fingerprints of biomolecules.[161, 162] With the large bandwidth and abundant spectral resources, THz band is increasingly recognized as a carrier frequency for the next-generation wireless communications. However, these promising applications still confront challenges from miniaturized and high-power sources, detectors with higher responsivities, modulators with low-power consumption and higher efficiency, switches with faster response and lower threshold, sensors with higher sensitivity and specificity, and isolators with larger isolation ratio and smaller footprint.[29, 163, 164] Efforts to improve the performance and functionality of THz devices using novel concepts, materials, and techniques have surged in recent years.[29, 165] The attractive physics of BICs and the emerging applications in shorter wavelengths would stimulate breakthroughs across all the bottlenecks in THz regime. Enhanced THz generation in LN thin films based on DFG processes has been demonstrated to achieve a 17-fold improvement by BICs compared with a bare thin film.[118] The significant enhancement of local pump energy would facilitate the miniaturization of THz sources for on-chip integration with low energy consumption. The topological features of BICs would enable the direct modulation of polarization and wavefront of THz emission, such as chiral and vortex radiation, which would benefit biosensing and multiplexing. Similar to generation of THz radiation, weak signals would be locally amplified by the strong field confinement of BICs, thereby improving the responsivities of detectors. The intrinsic polarization selectivity of BICs would exhibit a substantial discrimination ratio of spin light and provide solutions for filterless spin light detection.[94, 166]

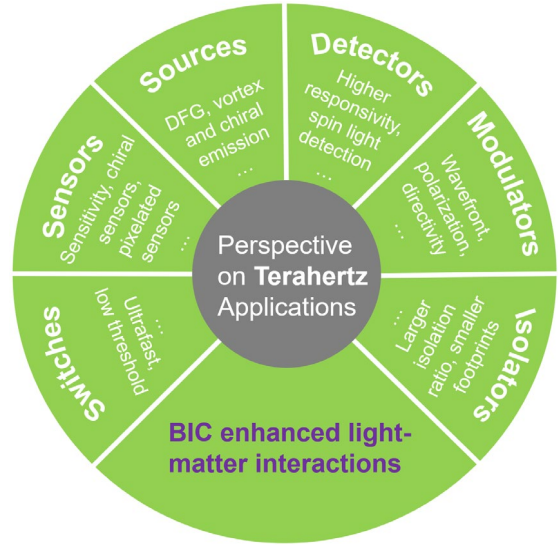

Figure. 9 Perspective of BICs on terahertz applications.

The amplified local field of BICs would lead to a higher sensitivity and contribute to the development of label-free sensors in THz applications. Deep subwavelength detection of refractive index perturbation has been demonstrated in the configuration of a flexible BIC metasurface.[167] The intrinsic chirality enabled by topological charge splitting and merging of BICs could maximize the circular dichroism of THz spectrum which would contribute to accurate characterization for pharmaceutical preparations and chiral molecules such as DNA and protein molecules.[89] More interesting studies may lead to the development of advanced sensing platforms with pixelated BIC metasurfaces and correlate the spatial information with BIC spectrum.

Inspired by the concept of nonlocal BIC metasurfaces, it would be intriguing to explore the manipulation of wavefront at specific frequency points while leaving the remaining bandwidth transparent. This approach has the potential to enable flexible shaping of THz pulses in both the spectral and spatial domains.[168, 169] Polarization states as well as spin-orbit coupling of THz waves could be modulated by BICs in dielectric metasurfaces with much higher efficiency than conventional metallic counterparts.[170, 171] The strategy to open a selected diffraction channel with 0-order channel ceased by BICs will benefit THz antennas for directional emission with higher directivity.

The enhanced light-matter interactions facilitated by BICs render them highly sensitive to external perturbations, making them useful not just for sensing applications, but also for ultrafast switches. Hybrid metasurfaces incorporating active media are commonly used as prototypes to demonstrate the ultrafast switches,[64-66, 172, 173] and the ultrasensitive nature of BICs will significantly lower the threshold for an efficient modulation depth.[174, 175] Ultrafast switches with low energy consumption are critical components for the next-generation THz communications.[158] Another essential component in communication systems is nonreciprocal devices, which usually rely on magneto-optical materials, leading to a large footprint of isolators or circulators. The enhanced light-matter interactions enabled by BICs could eliminate the rigid requirements of bulky magneto-optical materials or large external magnetic fields, and the smart manipulation of band dispersion of PhCs would help increase the isolation ratio.[176, 177]

The recent advances of BICs in theory and applications have demonstrated the important implications for engineering resonances in photonic devices. With the burgeoning deployment of photonics in industry, it is expected that BIC-enabled photonics will remain a very active research area and more critical progress would be achieved not only in classical optical regime but also in quantum photonics.

**Acknowledgments**

**General**: Thank others for any contributions.

**Author contributions:** L. Cong initiated the idea and supervised the project. G. Xu led the writing and editing with input from H. Xing, J. Fan, Z. Xue, D. Lu, and P. Shum. Subsection II "Engineering BICs from topological charges" was prepared by H. Xing and J. Fan. Subsection III "Applications" and subsection IV "Perspective on terahertz applications" were prepared by Z. Xue, D. Lu, J. Fan and G. Xu. All authors read and commented on the manuscript.

**Funding:** This work was supported by the National Natural Science Foundation of China (Award No.: 62175099), Guangdong Basic and Applied Basic Research Foundation (Award No.: 2023A1515011085), Stable Support Program for Higher Education Institutions from Shenzhen Science, Technology & Innovation Commission (Award No.: 20220815151149004), Global recruitment program of young experts of China, and startup funding of Southern University of Science and Technology.

**Conflicts of interest:** The authors declare no conflicts of interest.

**Data Availability:** The data are available from the corresponding author upon reasonable request.